\documentclass[twocolumn,pra,superscriptaddress,longbibliography]{revtex4-1}
\usepackage{graphicx,amsmath,amssymb}
\usepackage[colorlinks=true, citecolor=blue, urlcolor=blue, linkcolor=red]{hyperref}

\begin{document}
\title{Criticality-enhanced quantum sensor at finite temperature}

\author{Wei Wu}
\email{weiwu@lzu.edu.cn}
\affiliation{ Key Laboratory of Theoretical Physics of Gansu Province, \\
and Lanzhou Center for Theoretical Physics, Lanzhou University, Lanzhou, China}
\author{Chuan Shi}
\affiliation{ Key Laboratory of Theoretical Physics of Gansu Province, \\
and Lanzhou Center for Theoretical Physics, Lanzhou University, Lanzhou, China}

\begin{abstract}
Conventional criticality-based quantum metrological schemes work only at zero or very low temperature because the quantum uncertainty around the quantum phase transition point is generally erased by thermal fluctuations with the increase of temperature. Such an ultra-low temperature requirement severely restricts the development of quantum critical metrology. In this paper, we propose a thermodynamic-criticality-enhanced quantum sensing scenario at finite temperature. In our scheme, a qubit is employed as a quantum sensor to estimate parameters of interest in the Dicke model which experiences a thermodynamic phase transition. It is revealed that the thermodynamic criticality of the Dicke model can significantly improve the sensing precision. Enriching the scope of quantum critical metrology, our finding provides a possibility to realize highly sensitive quantum sensing without cooling.
\end{abstract}

\maketitle

\section{Introduction}

Quantum sensing concerns acquiring an ultra-high-precision estimation or measurement of a quantity of interest by making use of certain quantum resources~\cite{RevModPhys.89.035002}, which have no classical counterparts. The most common resources employed in quantum sensing are entanglement~\cite{RevModPhys.90.035005,PhysRevLett.121.160502,Nagata726,PhysRevLett.124.060402} and quantum squeezing~\cite{PhysRevLett.113.103004,PhysRevLett.119.193601,PhysRevLett.123.040402}. As reported in many previous articles~\cite{RevModPhys.90.035005,Nagata726,PhysRevLett.119.193601,PhysRevLett.123.040402}, these quantum resources can be employed to surpass the so-called shot-noise limit (SNL), which is a fundamental limit set by the law of classical statistics. Thus, quantum sensing provides an alternative framework for realizing a highly sensitive estimation or measurement of a parameter of interest and plays an important role in both theoretical and experimental studies. By far, quantum sensing has been widely applied to various domains, such as gravitational wave detection~\cite{PhysRevLett.123.231107,PhysRevLett.123.231108}, quantum radar~\cite{PhysRevLett.124.200503}, various quantum magnetometers~\cite{PhysRevX.10.011018,PhysRevA.99.062330} and quantum thermometries~\cite{PhysRevLett.114.220405,PhysRevA.98.050101}.

In Refs.~\cite{PhysRevA.80.012318,PhysRevA.78.042106,PhysRevA.78.042105,Wang_2014,PhysRevE.93.052118,PhysRevApplied.9.064006,PhysRevLett.124.120504,PhysRevA.103.023317}, the authors found the quantum Fisher information (FI) exhibits certain singular behaviors close to the quantum phase transition point of a many-body system. Due to the fact that the quantum FI represents the best attainable precision in quantum parameter estimation, these results mean the quantum criticality may be used to improve the sensitivity in quantum metrology. Such a criticality-based metrological scenario is called quantum critical metrology~\cite{PhysRevLett.121.020402,PhysRevLett.126.010502,PhysRevA.101.043609}, which commonly uses quantum uncertainties around the quantum critical point to improve the metrological precision. However, strictly speaking, the quantum phase transition happens only at zero temperature at which the thermal fluctuation is completely suppressed. In this sense, to take advantage of quantum criticality, the metrological protocol should be designed at very low temperatures, which should be as close to the absolute zero-temperature as possible. Such a requirement severely restricts the experimental realization of quantum-criticality-based metrology because no real installation works in the zero-temperature limit. An interesting question arises here: whether or not a thermodynamic phase transition, which happens at finite temperature, can be utilized to improve the quantum sensing performance.

\begin{figure}
\centering
\includegraphics[angle=0,width=4.5cm]{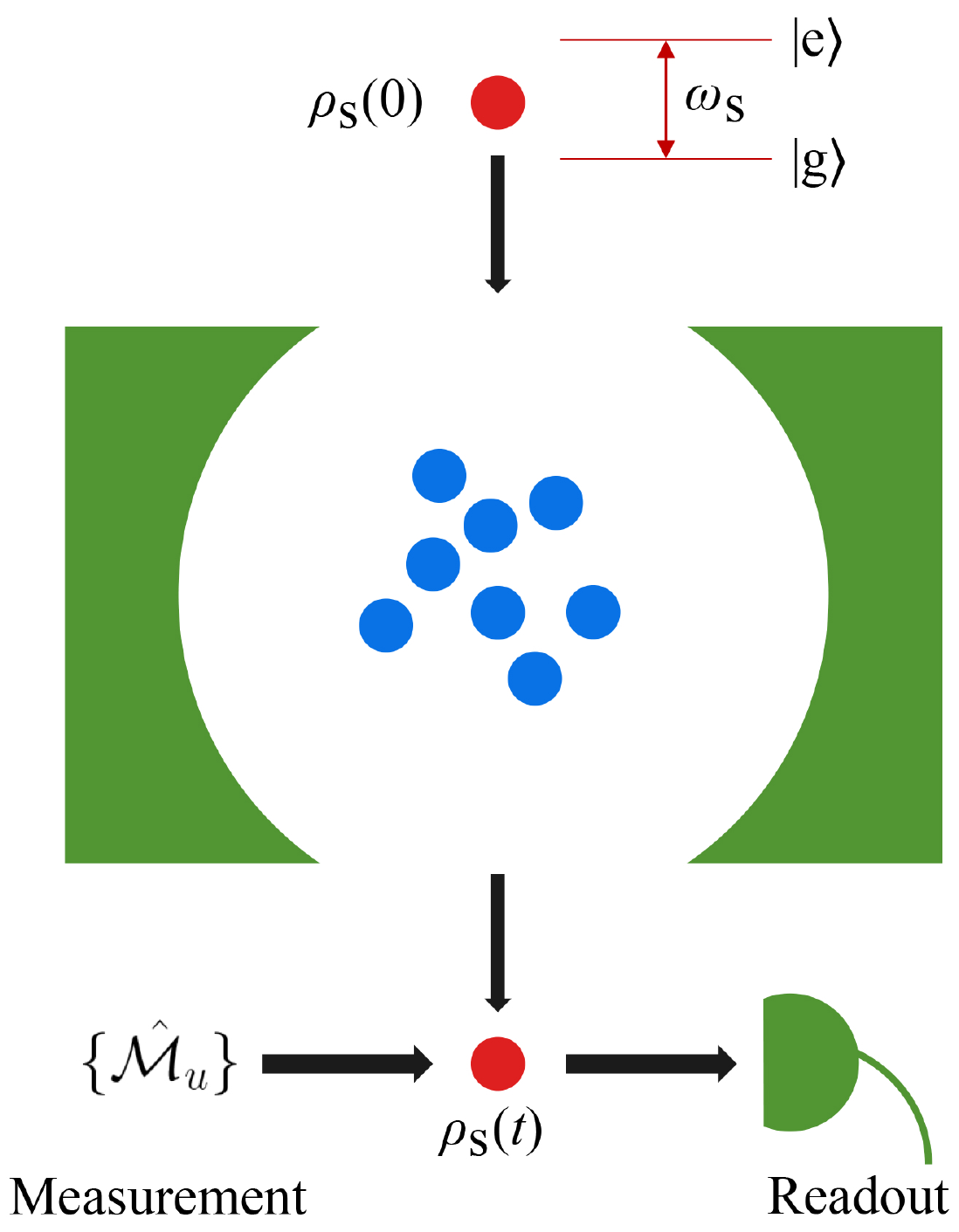}
\caption{Schematic diagram of our quantum sensing scheme in which a probe-qubit is used to detect the atom-cavity coupling strength in a DM Hamiltonian.}\label{fig:fig0}
\end{figure}

To address the above question, we propose a criticality-based quantum sensing scheme at finite temperature (see Fig.~\ref{fig:fig0}). In our scheme, a qubit, acting as a quantum sensor, is employed to detect the parameter of a Dicke model (DM)~\cite{PhysRev.93.99}, which experiences a thermodynamic phase transition at critical temperature~\cite{PhysRevA.9.418,PhysRevA.70.033808,Liberti2005,Bastarrachea_Magnani_2016,PhysRevE.96.012121}. It is revealed that the sensing precision can be significantly boosted close to the phase-transition point of the DM. Our result demonstrates the thermodynamic criticality can be used as a resource to improve the sensing performance and enrich the research of quantum critical metrology.

This paper is organized as follows. In Sec.~\ref{sec:sec1}, we first recall the thermodynamic characteristic of the well-known DM. In Sec.~\ref{sec:sec2}, we outline some basic concepts as well as the general formalism in quantum parameter estimation theory. In Sec.~\ref{sec:sec3}, we propose our sensing scheme and analyze its performance in detail. The main conclusions of this paper are drawn in Sec.~\ref{sec:sec4}. In several appendixes, we provide some additional details about the main text. Throughout the paper, we set $\hbar=k_{\mathrm{B}}=1$, and all the other units are dimensionless as well.

\section{Thermodynamic properties of the DM}\label{sec:sec1}

The Hamiltonian of the famous DM, which describes $N$ identical two-level atoms interacting with a single-mode cavity field, is given by
\begin{equation}\label{eq:eq1}
\hat{H}_{\mathrm{D}}=\epsilon\hat{J}_{z}+\omega\hat{a}^{\dagger}\hat{a}+\frac{g}{\sqrt{N}}(\hat{J}_{+}+\hat{J}_{-})(\hat{a}^{\dagger}+\hat{a}),
\end{equation}
where $\hat{J}_{z}=\frac{1}{2}\sum_{n=1}^{N}\hat{\sigma}_{n}^{z}$ and $\hat{J}_{\pm}=\sum_{n=1}^{N}\hat{\sigma}_{n}^{\pm}$, with $N$ being the number of the atoms, are collective operators of the atomic ensemble. The parameter $\epsilon$ denotes the transition frequency of the atomic ensemble's collective operator $\hat{J}_{z}$. Operators $\hat{a}^{\dagger}$ and $\hat{a}$ are creation and annihilation operators of the single-mode cavity field with corresponding frequency $\omega$, respectively. And the parameter $g$ quantifies the coupling strength between the atomic ensemble and the cavity field.

In the high-temperature regime, the partition function of the DM, i.e., $\mathcal{Z}_{\mathrm{D}}\equiv \mathrm{Tr}\exp(-\beta\hat{H}_{\mathrm{D}})$, can be analytically obtained (see Appendix A for details). With $\mathcal{Z}_{\mathrm{D}}$ at hand, one can easily see the free energy per atom in the thermodynamic limit is given by
\begin{equation}\label{eq:eq2}
f\equiv-\lim_{N\rightarrow\infty}\frac{1}{N}\frac{1}{\beta}\ln\mathcal{Z}_{\mathrm{D}}=\Phi(z_{0}),
\end{equation}
where $\Phi(z)$ is defined by
\begin{equation}\label{eq:eq3}
\Phi(z)\equiv-\beta\omega z^{2}+\ln\Bigg{[}2\cosh\bigg{(}\frac{\beta}{2}\sqrt{\epsilon^{2}+16g^{2}z^{2}}\bigg{)}\Bigg{]},
\end{equation}
and $z_{0}$ is determined by $\phi(z_{0})=0$, with $\phi(z)\equiv\partial_{z}\Phi(z)$. As shown in Ref.~\cite{PhysRevA.9.418}, there are two roots for the equation $\phi(z_{0})=0$, depending on the critical temperature $\beta_{c}$, which is given by
\begin{equation}\label{eq:eq4}
\beta_{c}=\frac{2}{\epsilon}\mathrm{arctanh}\Big{(}\frac{\epsilon\omega}{4g^{2}}\Big{)}.
\end{equation}
When $\beta<\beta_{c}$, the equation $\phi(z_{0})=0$ has a trivial solution, $z_{0}=0$, corresponding to the case in which the atomic ensemble and the cavity field are completely decoupled. On the other hand, if $\beta\geq\beta_{c}$, a nontrivial solution, $z_{0}=\sqrt{\epsilon^{2}\eta^{2}-\epsilon^{2}}/(4g)$, with $\eta$ determined by
\begin{equation}\label{eq:eq5}
\frac{\epsilon\omega}{4g^{2}}\eta=\tanh\Big{(}\frac{1}{2}\beta\eta\epsilon\Big{)},
\end{equation}
can be found. From the above analysis, a second-order thermodynamic phase transition occurs at the critical temperature $\beta=\beta_{c}$ provided $g>\frac{1}{2}\sqrt{\epsilon\omega}$~\cite{HEPP1973360,PhysRevA.7.831,PhysRevA.9.418,PhysRevA.70.033808,Liberti2005,Bastarrachea_Magnani_2016,PhysRevE.96.012121}. Above the critical temperature,
the DM is in the normal phase. However for $\beta\geq\beta_{c}$, the DM is in the superradiant phase~\cite{PhysRevA.9.418,PhysRevA.70.033808,Liberti2005,Bastarrachea_Magnani_2016,PhysRevE.96.012121}.

\begin{figure}
\centering
\includegraphics[angle=0,width=8.5cm]{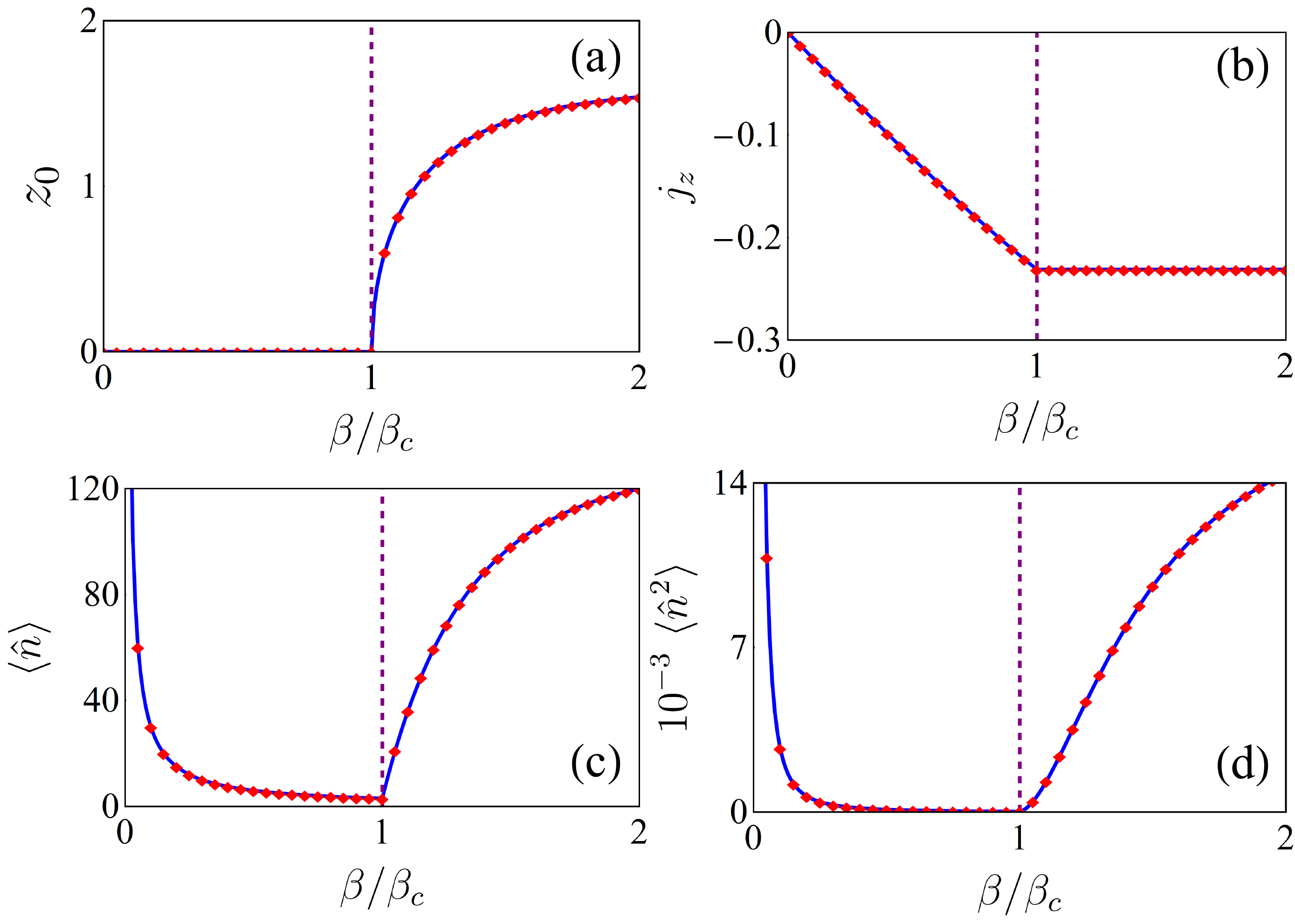}
\caption{(a) The value of $z_{0}$ versus the scaled temperature $\beta/\beta_{c}$. In the normal phase, we have $z_{0}=0$, while the occurrence of a nonzero value of $z_{0}$ in the regime $\beta/\beta_{c}>1$ means a phase transition. (b) The expected value of $\hat{J}_{z}$ per atom versus $\beta/\beta_{c}$. The expected values of (c) $\hat{n}$ and (d) $\hat{n}^{2}$ versus $\beta/\beta_{c}$ with $N=50$. The other parameters are chosen to be $\epsilon=1$, $g=0.3\epsilon$ and $\omega=4\tanh(\frac{1}{2}\epsilon)g^{2}\epsilon^{-1}$.}\label{fig:fig1}
\end{figure}

In Fig.~\ref{fig:fig1}, we plot the value of $z_{0}$ and expected value of $\hat{J}_{z}$ per atom, namely $j_{z}\equiv\langle \hat{J}_{z}\rangle/N$, as a function of $\beta/\beta_{c}$. One can see $j_{z}$ exhibits two different thermodynamic behaviors in the normal and super-radiant phases, implying $j_{z}$ is a good order parameter. Singular behavior of the first derivative of $j_{z}$ at $\beta/\beta_{c}=1$ means a thermodynamic phase transition.

\section{Quantum parameter estimation}\label{sec:sec2}

In this section, we would like to outline some basic concepts as well as some important formalism in quantum parameter estimation theory. Generally, to sense a physical quantity $\theta$ in a quantum system, one first needs to prepare a quantum sensor in an initial input state $\varrho_{\mathrm{in}}$ and couple the sensor to the system of interest. Due to the sensor-system interaction, the message of $\theta$ can be encoded into the output state of the sensor via $\varrho_{\mathrm{out}}=\Lambda_{\theta}(\varrho_{\mathrm{in}})=\varrho_{\theta}$, where $\Lambda_{\theta}$ is a $\theta$-dependent mapping. Commonly, $\Lambda_{\theta}$ can be realized by unitary~\cite{PhysRevLett.79.3865,PhysRevLett.124.010507,Chabuda2020} or nonunitary dynamics~\cite{PhysRevA.102.032607,PhysRevA.103.L010601,PhysRevApplied.15.054042}. Once the output state $\varrho_{\theta}$ is obtained, the information about $\theta$ can be extracted by measuring a certain sensor's observables.

In the above sensing scenario, the sensing precision corresponding to a given measurement scheme is constrained by the Cram$\mathrm{\acute{e}}$r-Rao bound $\delta\theta\geq 1/\sqrt{\upsilon F_{\theta}}$, where $\delta\theta$ is the standard error, $\upsilon$ is the number of repeated measurements (we set $\upsilon=1$ for the sake of convenience in this paper), and $F_{\theta}$ is the classical FI for the selected measurement scheme. In the language of quantum mechanics, to perform a quantum measurement, one needs to construct a set of positive-operator valued measurement operators $\{\mathcal{\hat{M}}_{u}\}$, which satisfy $\sum_{u}\hat{\mathcal{M}}_{u}^{\dagger}\hat{\mathcal{M}}_{u}=\mathbf{\hat{1}}$ with discrete measurement outcomes $\{u\}$. For an arbitrary output state $\varrho_{\mathrm{\theta}}$, operator $\mathcal{\hat{M}}_{u}$ yields an outcome $u$ with corresponding probability distribution $p(u|\theta)=\mathrm{Tr}(\mathcal{\hat{M}}_{u}\varrho_{\theta}\mathcal{\hat{M}}_{u}^{\dagger})$. With all the probabilities from $\{\mathcal{\hat{M}}_{u}\}$ at hand, the classical FI can be computed as~\cite{SMKay,Zhang_2018}
\begin{equation}\label{eq:eq6}
F_{\theta}=\sum_{u}p(u|\theta)\bigg{[}\frac{\partial}{\partial\theta}\ln p(u|\theta)\bigg{]}^{2}.
\end{equation}

From Eq.~(\ref{eq:eq6}), one can find the value of classical FI strongly relies on the form of measurement operators $\{\mathcal{\hat{M}}_{u}\}$. Running over all the possible measurements, the ultimate sensing precision of $\theta$ is then bounded by the quantum Cram$\mathrm{\acute{e}}$r-Rao inequality $\delta\theta\geq 1/\sqrt{\upsilon \mathcal{F}_{\theta}}$, where $\mathcal{F}_{\theta}\equiv \mathrm{Tr}(\hat{\varsigma}^{2}\varrho_{\theta})$ with $\hat{\varsigma}$ determined by $\partial_{\theta}\varrho_{\theta}=\frac{1}{2}(\hat{\varsigma}\varrho_{\theta}+\varrho_{\theta}\hat{\varsigma})$ is the quantum FI. Specially, if the output state $\varrho_{\theta}$ is a two-dimensional density matrix described in the Bloch representation, namely, $\varrho_{\theta}=\frac{1}{2}(\mathbf{1}_{2}+\pmb{r}\cdot\pmb{\hat{\sigma}})$ with $\pmb{r}$ being the Bloch vector and $\pmb{\hat{\sigma}}\equiv(\hat{\sigma}_{x},\hat{\sigma}_{y},\hat{\sigma}_{z})$ being the vector of Pauli matrices, the quantum FI can easily be calculated via the following corollary~\cite{Liu_2019}:
\begin{equation}\label{eq:eq7}
\mathcal{F}_{\theta}=|\partial_{\theta}\pmb{r}|^{2}+\frac{(\pmb{r}\cdot\partial_{\theta}\pmb{r})^{2}}{1-|\pmb{r}|^{2}}.
\end{equation}
For the pure state case, the above equation further reduces to $\mathcal{F}_{\theta}=|\partial_{\theta}\pmb{r}|^{2}$.

Physically speaking, $\mathcal{F}_{\theta}$ describes all the statistical information contained in the output state $\varrho_{\theta}$, while $F_{\theta}$ describes the statistical message extracted by projecting the selected measurement operators $\{\mathcal{\hat{M}}_{u}\}$ onto $\varrho_{\theta}$. In this sense, one can immediately conclude that $\mathcal{F}_{\theta}\geq F_{\theta}$. As long as the selected measurement scheme is the optimal one, $F_{\theta}$ can be saturated to $\mathcal{F}_{\theta}$. Unfortunately, the optimal measurement scheme is generally difficult to obtain. How to design a quantum measurement scheme saturating the best attainable precision determined by the quantum FI is important in quantum sensing.

\section{Our sensing scheme}\label{sec:sec3}

\subsection{Single-qubit case}

As displayed in Fig.~\ref{fig:fig0}, we use a qubit as the quantum sensor to measure the coupling strength $g$ of the DM Hamiltonian. To this aim, a probe qubit is injected into the single-mode cavity field and interacts with the cavity accordingly~\cite{PhysRevA.80.063829,PhysRevA.87.024101,Wu2016}. Assuming the quantum sensor and the cavity field are far-off-resonant and the coupling between them is weak, an effective Hamiltonian of the whole sensor-DM system can be described by (see Appendix B for details)
\begin{equation}\label{eq:eq8}
\hat{H}_{\mathrm{eff}}=\omega_{\mathrm{s}}\hat{\sigma}_{+}\hat{\sigma}_{-}+\hat{H}_{\mathrm{D}}+\lambda\hat{\sigma}_{z}\hat{a}^{\dagger}\hat{a},
\end{equation}
where $\omega_{\mathrm{s}}$ is the effective frequency of the qubit and $\lambda$ denotes the effective coupling strength between the qubit and the single-mode cavity field.

In this section, we assume the initial state of the whole sensor-DM system is given by $\rho_{\mathrm{tot}}(0)=|\psi_{\mathrm{s}}(0)\rangle\langle\psi_{\mathrm{s}}(0)|\otimes\rho_{\mathrm{D}}$, where $|\psi_{\mathrm{s}}(0)\rangle=(|\mathrm{e}\rangle+|\mathrm{g}\rangle)/\sqrt{2}$ with $|\mathrm{e},\mathrm{g}\rangle$ being the eigenstates of $\hat{\sigma}_{z}$ and $\rho_{\mathrm{D}}$ being the thermal Gibbs state of the DM. Then, the reduced dynamics of the sensor can be exactly obtained from the quantum von Neumann equation $\partial_{t}\rho_{\mathrm{tot}}(t)=-i[\hat{H}_{\mathrm{eff}},\rho_{\mathrm{tot}}(t)]$. After tracing out the degree of freedom of the DM Hamiltonian, we find the reduced density operator of the sensor in the basis $\{|\mathrm{e}\rangle,|\mathrm{g}\rangle\}$ reads
\begin{equation}\label{eq:eq9}
\rho_{\mathrm{s}}(t)=\left[
                       \begin{array}{cc}
                         \rho_{\mathrm{ee}}(0) & \rho_{\mathrm{eg}}(0)\mathcal{L}(t) \\
                         \rho_{\mathrm{ge}}(0)\mathcal{L}^{*}(t) & \rho_{\mathrm{gg}}(0) \\
                       \end{array}
                     \right],
\end{equation}
where $\rho_{\mathrm{ii}}(0)\equiv\langle \mathrm{i}|\rho_{\mathrm{s}}(0)|\mathrm{i}\rangle=1/2$ with $\mathrm{i}=\mathrm{e},\mathrm{g}$ and the decoherence factor $\mathcal{L}(t)$ is given by
\begin{equation}\label{eq:eq10}
\mathcal{L}(t)=\mathrm{Tr}_{\mathrm{D}}\Big{(}e^{-it\hat{H}_{\mathrm{e}}}\hat{\rho}_{\mathrm{D}}e^{it\hat{H}_{\mathrm{g}}}\Big{)},
\end{equation}
with $\hat{H}_{\mathrm{e}}=\hat{H}_{\mathrm{D}}+\lambda \hat{a}^{\dagger}\hat{a}+\omega_{\mathrm{s}}$ and $\hat{H}_{\mathrm{g}}=\hat{H}_{\mathrm{D}}-\lambda \hat{a}^{\dagger}\hat{a}$. In the weak-coupling regime, i.e., $\lambda/\omega_{\mathrm{s}}\ll 1$, one can find (see Appendix C for details)
\begin{equation}\label{eq:eq11}
\mathcal{L}(t)\simeq e^{-it\omega_{\mathrm{s}}}e^{-2i\lambda t\langle\hat{n}\rangle}e^{-\lambda^{2}t^{2}\langle\hat{n}^{2}\rangle},
\end{equation}
where $\hat{n}\equiv\hat{a}^{\dagger}\hat{a}$ is the photon operator. The expressions of $\langle \hat{n}\rangle$ and $\langle \hat{n}^{2}\rangle$ are given by (see Appendix A)
\begin{equation}\label{eq:eq12}
\langle \hat{n}\rangle=\frac{1}{2\beta\omega }+N z_{0}^{2},
\end{equation}
\begin{equation}\label{eq:eq13}
\langle \hat{n}^{2}\rangle=\frac{3}{4\beta^{2}\omega^{2} }+\frac{N}{\beta\omega }z_{0}^{2}+N^{2}z_{0}^{4}.
\end{equation}
In Fig.~\ref{fig:fig1}, we plot $\langle \hat{n}\rangle$ and $\langle \hat{n}^{2}\rangle$ as a function of $\beta/\beta_{c}$. It is clear that $\langle \hat{n}\rangle$ and $\langle \hat{n}^{2}\rangle$ experience sudden changes at the critical point $\beta/\beta_{c}=1$. This result means the decoherence factor will be sensitive to the thermodynamic properties of the DM. In this sense, we expect the sensing performance also exhibits a singular behavior near the thermodynamic critical point.
\begin{figure}
\centering
\includegraphics[angle=0,width=8.5cm]{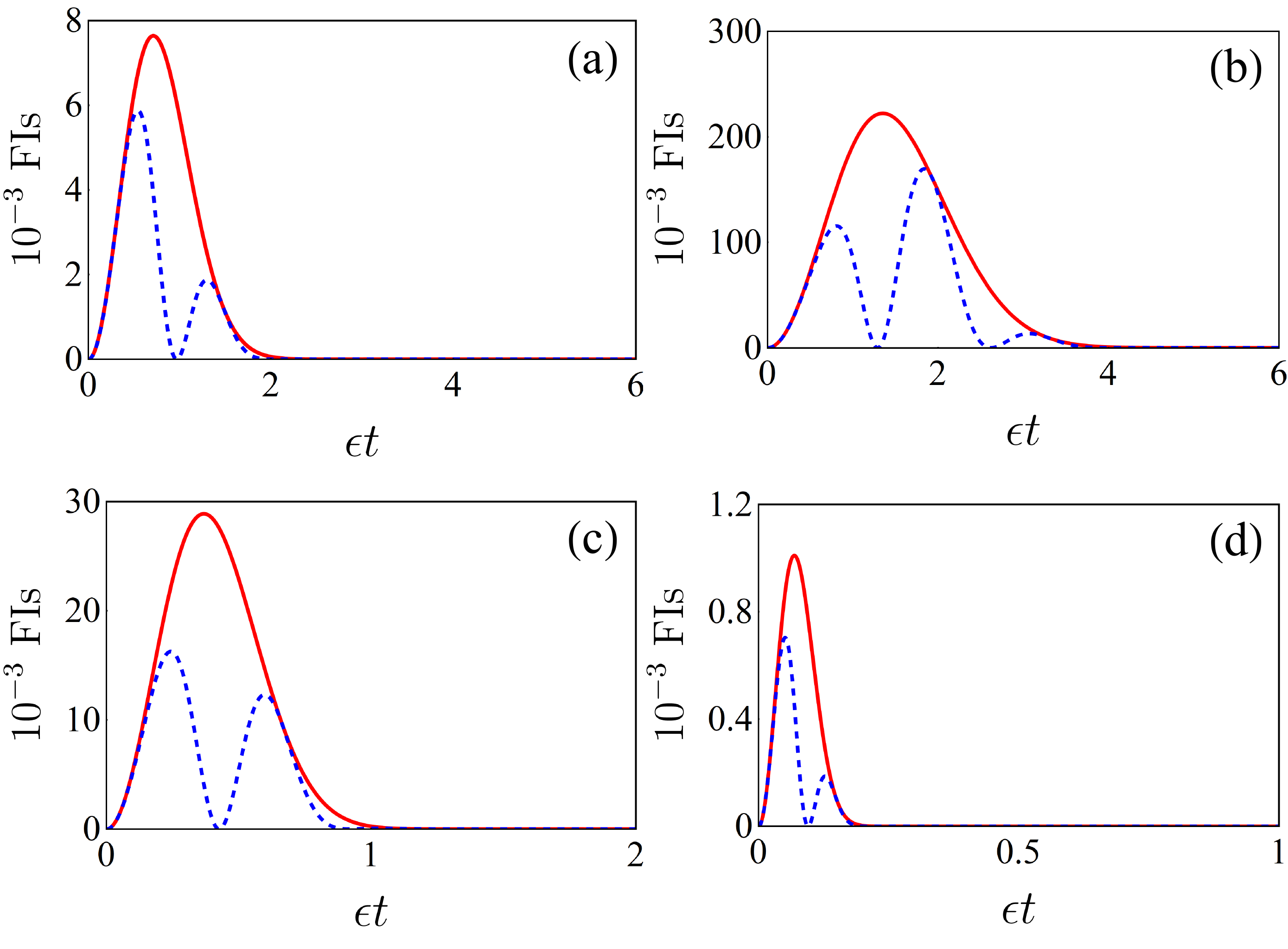}
\caption{ The dynamics of classical FI $F_{g}(t)$ (blue dashed lines) and quantum FI $\mathcal{F}_{g}(t)$ (red solid lines) with $\lambda=0.1\epsilon$ and $\omega_{\mathrm{s}}=1.5\epsilon$ for different temperatures: (a)$\beta/\beta_{c}=0.5$ ,(b) $\beta/\beta_{c}=0.95$, (c)$\beta/\beta_{c}=1.05$, and (d)$\beta/\beta_{c}=1.5$. Other parameters the same with these of Fig.~\ref{fig:fig1}.}\label{fig:fig2}
\end{figure}
\begin{figure}
\centering
\includegraphics[angle=0,width=8.5cm]{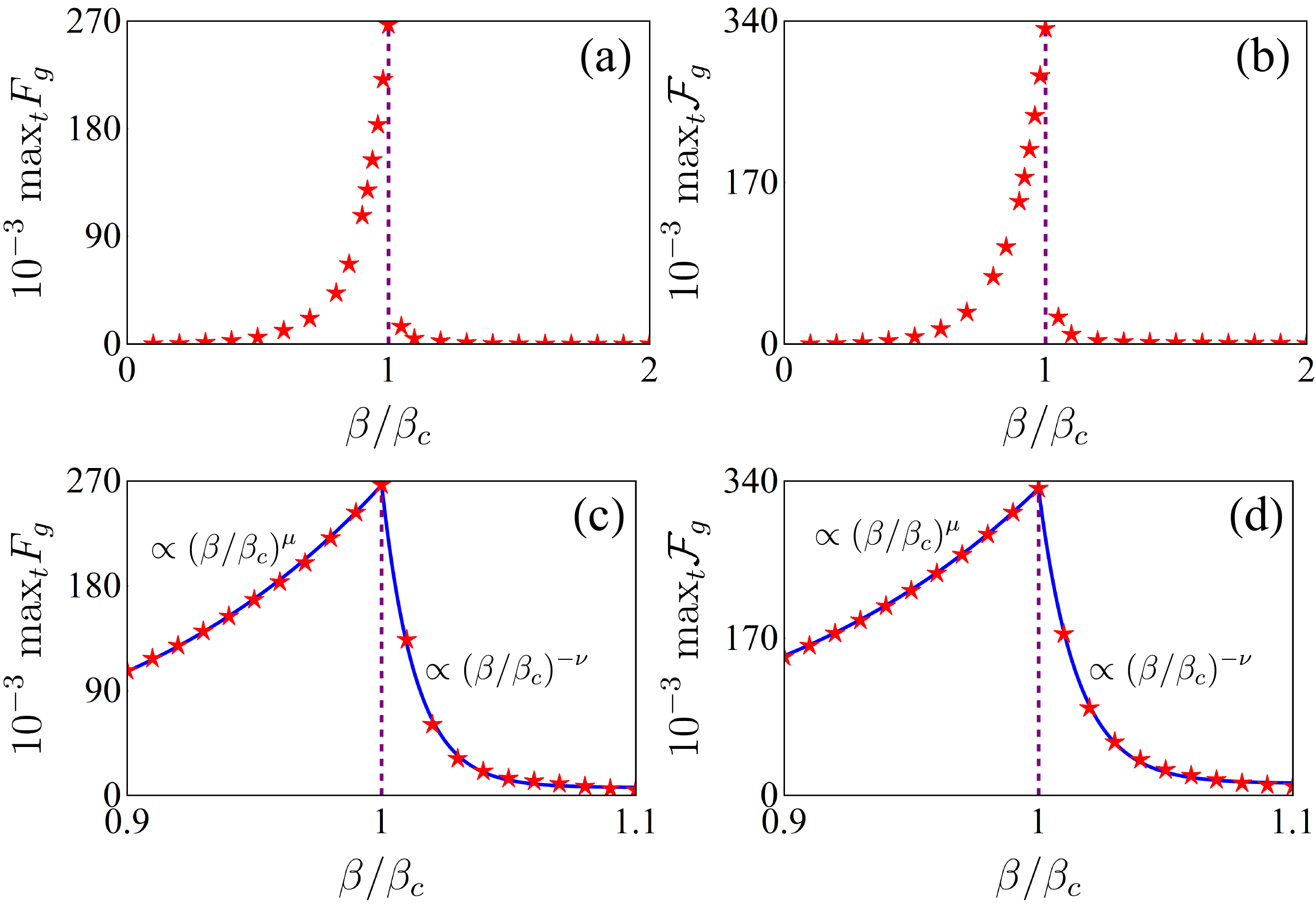}
\caption{By optimizing the encoding time, the corresponding maximum classical and quantum FIs are plotted as a function of $\beta/\beta_{c}$. Red stars are from numerical calculations, and blue solid lines are obtained using the least-squares fittings. In the vicinity of the critical region, we find, in the norm phase, the FIs behave as $\mathrm{F}\propto (\beta/\beta_{c})^{\mu}$ where (c)$\mu\simeq 9.5$  and (d)$\nu\simeq 9.1$, while in the super-radiant phase, $\mathrm{F}\propto(\beta/\beta_{c})^{-\nu}$, where (c) $\nu\simeq 76$ and (d) $\nu\simeq 67$. Other parameters are the same as those in Fig.~\ref{fig:fig1}.}\label{fig:fig3}
\end{figure}

With the help of $\rho_{\mathrm{s}}(t)$, the expressions for FIs can be immediately obtained with the help of Eqs.~(\ref{eq:eq6}) and ~(\ref{eq:eq7}). The selected measurement operators $\{\mathcal{\hat{M}}_{u}\}$ in our scheme, the explicit expressions of classical and quantum FIs are given in Appendix D. Although the measurement scheme selected can not saturate the upper bound given by the quantum Cram$\mathrm{\acute{e}}$r-Rao theorem, the corresponding classical FI is in qualitative agreement with the quantum FI. In Fig.~\ref{fig:fig2}, we display the dynamics of classical and quantum FIs. One can see the FIs gradually increase from their initial values $\mathrm{F}(0)=0$ to the maximum values. Then, the FIs begin to decrease and eventually disappear in the long-encoding-time limit. This result is physically reasonable because the whole sensing process can be classified into the following three steps. First, no message about $g$ is contained in the initial state of the sensor, resulting in $\mathrm{F}(0)=0$ at the beginning. Then FIs gradually increase with the evolution of encoding time, which is caused by the sensor-DM interaction generating the information about $g$ in $\rho_{\mathrm{s}}(t)$. Finally, in the long-encoding-time regime, the decoherence induced by the weak sensor-DM coupling leads to a trivial canonical thermal equilibrium state of the sensor as $\rho_{\mathrm{s}}(\infty)\propto \exp(-\beta\omega_{\mathrm{s}}\hat{\sigma}_{+}\hat{\sigma}_{-})$~\cite{doi:10.1063/1.4722336,PhysRevE.90.022122,doi:10.1063/1.5141519}, which is independent of the parameter $g$ and implies no information can be extracted, namely $\mathrm{F}(\infty)=0$. It is worth mentioning that the noncanonical characteristic occurs if the coupling is very strong~\cite{doi:10.1063/1.4722336,PhysRevE.90.022122}. In such circumstances, the long-time steady state may provide certain contributions to the FIs~\cite{PhysRevA.103.L010601}.

From the above analysis, an optimal encoding time exists to maximize the value of FI. This result can be physically understood as the competition (or interplay) between the indispensable sensor-DM interaction for encoding and the deterioration of quantum coherence induced by the nonunitary dynamics~\cite{PhysRevA.102.032607,PhysRevA.103.L010601}. In Fig.~\ref{fig:fig3}, we display the maximum FI with respect to the optimal encoding time as a function of $\beta/\beta_{c}$. In the vicinity of the critical region, we further find the FIs increase in the form of $\mathrm{F}\propto(\beta/\beta_{c})^{\mu}$ as $\beta$ approaches $\beta_{c}$ in the normal phase; however, in the super-radiant phase, the FIs exhibit certain power-law behaviors as $\mathrm{F}\propto(\beta/\beta_{c})^{-\nu}$ with $\mu\neq\nu$. A peak at the thermodynamic critical point $\beta/\beta_{c}=1$ is clearly revealed, which suggests the thermodynamic criticality can significantly improve the sensing precision at finite temperature. On the other hand, such singular behavior of the FI can be viewed as a signature of the second-order thermodynamic criticality of the DM~\cite{HEPP1973360,PhysRevA.7.831}. Additionally, we also extend our scheme to the multi-parameter sensing case in Appendix E, which displays a quite similar result. These results demonstrate the feasibility of our thermodynamic-criticality-enhanced quantum sensing scenario.

\subsection{Multi-qubit case}

Next, we generalize the above sensing scheme to a more general situation in which $\mathbb{N}$ identical qubit-probes are employed. We assume each sensor independently interacts with its own DM Hamiltonian. In this sense, the number of the qubits can be viewed as a quantum resource. We would like to study the relationship between the sensing precision and $\mathbb{N}$. In this section, three different initial states are taken into consideration: (i) an uncorrelated product state
\begin{equation}
|\psi_{\mathrm{s}}^{\mathrm{unc}}(0)\rangle=\frac{1}{\sqrt{2}}\Big{(}|\mathrm{e}\rangle+|\mathrm{g}\rangle\Big{)}^{\otimes\mathbb{N}},
\end{equation}
(ii) a Greenberger-Horne-Zeilinger(GHZ)-type maximally entangled state
\begin{equation}
|\psi_{\mathrm{s}}^{\mathrm{GHZ}}(0)\rangle=\frac{1}{\sqrt{2}}\Big{(}|\mathrm{e}\rangle^{\otimes\mathbb{N}}+|\mathrm{g}\rangle^{\otimes\mathbb{N}}\Big{)},
\end{equation}
and (iii) an $\mathbb{N}$-qubit Werner state~\cite{PhysRevA.40.4277,Eltschka_2014}, i.e.,
\begin{equation}
\rho_{\mathrm{s}}^{\mathrm{W}}(0)=\frac{w}{2^{\mathbb{N}}}\mathbf{\hat{1}}_{\mathbb{N}}+(1-w)|\psi_{\mathrm{s}}^{\mathrm{GHZ}}(0)\rangle\langle\psi_{\mathrm{s}}^{\mathrm{GHZ}}(0)|,
 \end{equation}
where $\mathbf{\hat{1}}_{\mathbb{N}}$ is a $2^{\mathbb{N}}\times2^{\mathbb{N}}$ identity matrix and the admixture parameter $w$, varying from 0 to 1, can be used to describe the strength of white noise~\cite{Eltschka_2014,PhysRevA.90.062113,Micadei_2015}. The Werner state $\rho_{\mathrm{s}}^{\mathrm{W}}(0)$ is particularly interesting because it is a mixture of a maximally entangled state ($w=0$) and a completely mixed state ($w=1$). By comparing the output results from the above three initial states, the effect of entanglement on the performance of our quantum sensing scheme can be explored.

\begin{figure}
\centering
\includegraphics[angle=0,width=7cm]{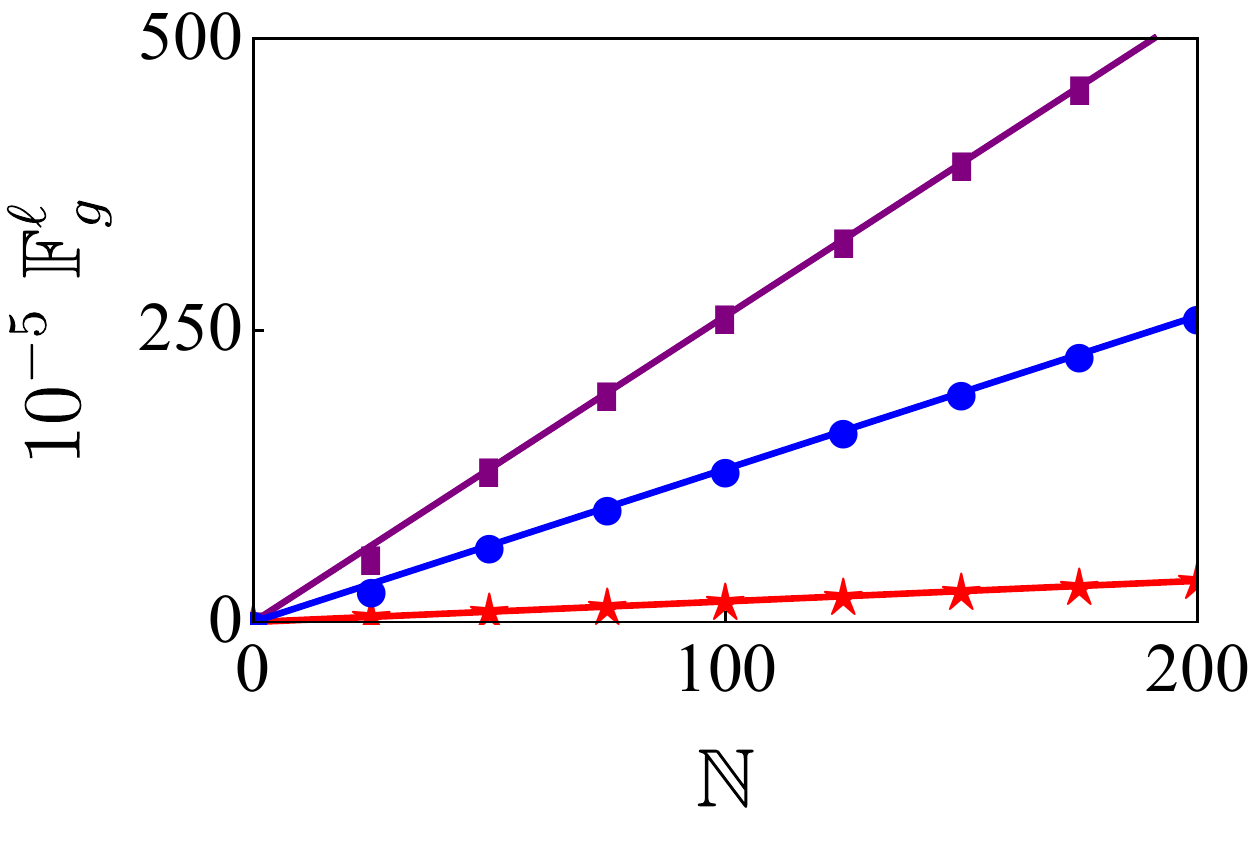}
\caption{By optimizing the encoding time and setting $\beta/\beta_{c}=1$, the optimal quantum FIs,$\mathbb{F}_{g}^{\ell}$ are plotted as a function of $\mathbb{N}$ with $\lambda=10^{-3}\epsilon$ and $w=0.5$: $\ell=\mathrm{unc}$ (red stars), $\ell=\mathrm{W}$ (blue circles), and $\ell=\mathrm{GHZ}$ (purple rectangles) are from numerical calculations. The solid lines are obtained by the least-squares fittings with the forms of $\mathbb{F}_{g}^{\ell}=\mathrm{const}.\times\mathbb{N}$. Other parameters are the same as those in Fig.~\ref{fig:fig1}.}\label{fig:fig4}
\end{figure}

For the uncorrelated state case, one can find the output state is just an $\mathbb{N}$-fold tensor product of $\rho_{\mathrm{s}}(t)$, namely, $\rho_{\mathrm{s}}^{\mathrm{unc}}(t)=[\rho_{\mathrm{s}}(t)]^{\otimes\mathbb{N}}$. Thus, owing to the additivity of quantum FI~\cite{Liu_2019}, we have $\mathcal{F}_{g}^{\mathrm{unc}}(t)=\mathbb{N}\mathcal{F}_{g}(t)$. However, for the GHZ-state case, one can find~\cite{PhysRevA.92.010302}
\begin{equation}\label{eq:eq14}
\begin{split}
\rho_{\mathrm{s}}^{\mathrm{GHZ}}(t)=&\frac{1}{2}\Big{(}|\pmb{\mathrm{e}}\rangle\langle\pmb{\mathrm{e}}|+|\pmb{\mathrm{g}}\rangle\langle\pmb{\mathrm{g}}|\\
&+e^{\mathbb{N}\ln\mathcal{L}(t)}|\pmb{\mathrm{e}}\rangle\langle\pmb{\mathrm{g}}|+\mathrm{H}.\mathrm{c}.\Big{)},
\end{split}
\end{equation}
which is still a two-dimensional matrix in the basis $\{|\pmb{\mathrm{e}}\rangle,|\pmb{\mathrm{g}}\rangle\}$ with $|\pmb{\mathrm{i}}\rangle\equiv|\mathrm{i}\rangle^{\otimes\mathbb{N}}$. Thus, the QFI for the GHZ-state case can also be calculated by using Eq.~(\ref{eq:eq7}). The explicit expression of $\mathcal{F}_{g}^{\mathrm{GHZ}}(t)$ is given in Appendix D. For the Werner state case, one can easily find
\begin{equation}
\rho_{\mathrm{s}}^{\mathrm{W}}(t)=\frac{w}{2^{\mathbb{N}}}\mathbf{\hat{1}}_{\mathbb{N}}+(1-w)\rho_{\mathrm{s}}^{\mathrm{GHZ}}(t).
\end{equation}
To calculate the quantum FI from $\rho_{\mathrm{s}}^{\mathrm{W}}(t)$, one needs to rearrange the above expression as $\rho_{\mathrm{s}}^{\mathrm{W}}(t)=\rho_{\mathrm{s},1}^{\mathrm{W}}(t)\oplus\rho_{\mathrm{s},2}^{\mathrm{W}}(t)$, where
\begin{equation}
\begin{split}
\rho_{\mathrm{s},1}^{\mathrm{W}}(t)=&\bigg{[}\frac{w}{2^{\mathbb{N}}}+\frac{1}{2}(1-w)\bigg{]}(|\pmb{\mathrm{e}}\rangle\langle\pmb{\mathrm{e}}|+|\pmb{\mathrm{g}}\rangle\langle\pmb{\mathrm{g}}|)\\
&+\bigg{[}\frac{1}{2}(1-w)e^{\mathbb{N}\ln\mathcal{L}(t)}|\pmb{\mathrm{e}}\rangle\langle\pmb{\mathrm{g}}|+\mathrm{H}.\mathrm{c}.\bigg{]},
\end{split}
\end{equation}
and
\begin{equation}
\begin{split}
\rho_{\mathrm{s},2}^{\mathrm{W}}(t)=\frac{w}{2^{\mathbb{N}}}\sum_{\mathrm{j}=1}^{\mathbb{N}-1}\mathbb{P}\bigg{[}\big{(}|\mathrm{e}\rangle\langle\mathrm{e}|\big{)}^{\otimes\mathrm{j}}\otimes\big{(}|\mathrm{g}\rangle\langle\mathrm{g}|\big{)}^{\otimes(\mathbb{N}-\mathrm{j}-1)}\bigg{]}.
\end{split}
\end{equation}
Here, $\mathbb{P}$ represents all possible permutations. Taking $\mathbb{N}=4$ and $\mathrm{j}=1$ as an example, the effect of $\mathbb{P}$ is
\begin{equation}
\begin{split}
\mathbb{P}\Big{(}|\mathrm{e}\rangle\langle\mathrm{e}|\otimes|\mathrm{gg}\rangle\langle\mathrm{gg}|\Big{)}&=|\mathrm{egg}\rangle\langle\mathrm{egg}|\\
&+|\mathrm{geg}\rangle\langle\mathrm{geg}|+|\mathrm{gge}\rangle\langle\mathrm{gge}|.
\end{split}
\end{equation}
Then, using the fact that the quantum FI is additive for a direct-sum density matrix, namely~\cite{Liu_2019},
\begin{equation}
\mathcal{F}\big{[}\varrho_{1}\oplus\varrho_{2}\big{]}= \mathcal{F}[\varrho_{1}]+\mathcal{F}[\varrho_{2}],
\end{equation}
where $\mathcal{F}[\varrho]$ denote the quantum FI with respect to $\varrho$, one can immediately find $\mathcal{F}_{g}^{\mathrm{W}}(t)=\mathcal{F}_{g}[\rho_{\mathrm{s},1}^{\mathrm{W}}(t)]$. Thus, the remaining task is quite clear: we turn to evaluate $\mathcal{F}_{g}[\rho_{\mathrm{s},1}^{\mathrm{W}}(t)]$. Because $\mathrm{Tr}[\rho_{\mathrm{s},1}^{\mathrm{W}}(t)]\neq 1$ in the basis $\{|\pmb{\mathrm{e}}\rangle,|\pmb{\mathrm{g}}\rangle\}$, $\rho_{\mathrm{s},1}^{\mathrm{W}}(t)$ can not be expressed in the standard Bloch representation. Thus, Eq.~(\ref{eq:eq7}) is no longer available. In this situation, to find $\mathcal{F}_{g}[\rho_{\mathrm{s},1}^{\mathrm{W}}(t)]$, one needs to diagonalize $\rho_{\mathrm{s},1}^{\mathrm{W}}(t)$ as $\rho_{\mathrm{s},1}^{\mathrm{W}}(t)=\sum_{l}\pi_{l}|\pi_{l}\rangle\langle\pi_{l}|$;\ then, $\mathcal{F}_{g}[\rho_{\mathrm{s},1}^{\mathrm{W}}(t)]$ can be calculated via the following theorem~\cite{Liu_2019}:
\begin{equation}
\begin{split}
\mathcal{F}_{g}[\rho_{\mathrm{s},1}^{\mathrm{W}}(t)]=&\sum_{l}\frac{(\partial_{g}\pi_{l})^{2}}{\pi_{l}}+\sum_{l}4\pi_{l}\langle\partial_{g}\pi_{l}|\partial_{g}\pi_{l}\rangle\\
&-\sum_{l,l'}\frac{8\pi_{l}\pi_{l'}}{\pi_{l}+\pi_{l'}}|\langle\partial_{g}\pi_{l}|\pi_{l'}\rangle|^{2}.
\end{split}
\end{equation}
The concrete expression for $\mathcal{F}_{g}[\rho_{\mathrm{s},1}^{\mathrm{W}}(t)]$ can be exactly obtained from the above equation, but it is rather complicated, leading a counter-intuitive physical picture. Fortunately, in the limit $\mathbb{N}\gg1$, one can find $\rho_{\mathrm{s},1}^{\mathrm{W}}(t)$ reduces to $(1-w)\rho_{\mathrm{s}}^{\mathrm{GHZ}}(t)$. A physically visual result can be obtained
\begin{equation}
\begin{split}
\mathcal{F}_{g}^{\mathrm{W}}(t)=&\mathcal{F}_{g}[\rho_{\mathrm{s},1}^{\mathrm{W}}(t)]\\
\overset{\mathbb{N}\gg1}{=}&\mathcal{F}_{g}[(1-w)\rho_{\mathrm{s}}^{\mathrm{GHZ}}(t)]\\
=&(1-w)\mathcal{F}_{g}^{\mathrm{GHZ}}(t).
\end{split}
\end{equation}
This result satisfies the convexity of quantum FI for a Werner-like state~\cite{Liu_2019,PhysRevA.90.062113,Micadei_2015}. And one can immediately conclude that $\mathcal{F}_{g}^{\mathrm{GHZ}}(t)\geq\mathcal{F}_{g}^{\mathrm{W}}(t)$, implying that a maximally entangled initial state provides a better sensing performance than that of the Werner state.

Like for the single-qubit case, one can find $\max_{t}\mathcal{F}_{g}^{\ell}$, with $\ell=\mathrm{unc}$, $\mathrm{GHZ}$ and $\mathrm{W}$, reach their maximum values at the the thermodynamic critical point $\beta/\beta_{c}=1$. Thus, we can define a quantity
\begin{equation}\label{eq:eq15}
\mathbb{F}_{g}^{\ell}\equiv\max_{\{t,\beta\}}\mathcal{F}_{g}^{\ell}
\end{equation}
and view it as the ultimate precision in the multi-qubit sensing scheme. In Fig.~\ref{fig:fig4}, we plot $\mathbb{F}_{g}^{\ell}$ as the function of $\mathbb{N}$. One can observe $\mathbb{F}_{g}^{\ell}$ is proportional to the number of qubits; that is, $\mathbb{F}_{g}^{\ell}$ scales as SNL. Moreover, from Fig.~\ref{fig:fig4}, we find $\mathbb{F}_{g}^{\mathrm{GHZ}}>\mathbb{F}_{g}^{\mathrm{W}}>\mathbb{F}_{g}^{\mathrm{unc}}$, which means entanglement can be used as a quantum resource to improve the sensing precision.

\section{Conclusion}\label{sec:sec4}

In summary, we used a qubit-probe to reveal the atom-cavity coupling strength of the DM Hamiltonian which experiences a thermodynamic phase transition at the critical temperature. It is found that the FI with respect to the reduced density operator of the qubit exhibits a peak at the thermodynamic phase transition point of the DM. This result means the thermodynamic criticality can be viewed as a resource to boost the performance of quantum sensing. And vice versa, the singular behavior of FI can be utilized as a tool to reveal the second-order thermodynamic criticality of the DM. In comparison to conventional quantum-criticality-based metrological schemes, which work only at zero or very low temperature, our scheme provides a possibility to realize  highly sensitive quantum sensing at finite temperature. Moreover, our sensing scheme can be generalized to a case in which multiple qubits are employed. By increasing the number of qubits, the quantum FI linearly increases in the form of SNL. By comparing the sensing results from the uncorrelated and entangled input states, we also demonstrated the entanglement can effectively improve the sensing precision.

It is necessary to emphasize that our sensing scheme is totally different from many previous studies~\cite{PhysRevLett.121.020402,PhysRevLett.126.010502,PhysRevA.101.043609} in the following two aspects. On the one hand, we employed the thermodynamic phase transition, instead of the quantum phase transition, as the resource to boost the sensing performance. Our key recipe is the thermal fluctuations at the critical point, which lead to a dramatic change in the decoherence factor as well as a peak in FIs. In this sense, we expect our finding can be generalized to other many-body systems which exhibit thermodynamic phase transitions without difficulty. On the other hand, we used a non-unitary encoding dynamics, instead of the unitary encoding process, to realize our sensing scheme. Such non-unitarity stems from the indispensable sensor-DM coupling and can lead to a totally different result in comparison with that of the unitary case~\cite{PhysRevA.103.L010601}. Finally, we expect our results presented in this paper to be of interest for a wide range of experimental applications in quantum sensing and quantum metrology.

\section{Acknowledgments}\label{sec:sec5}

W. Wu wishes to thank Dr. Zi-Ling Luo, Dr. Si-Yuan Bai, Dr. Chong Chen, Dr. Jing Liu and Prof. Jun-Hong An for many fruitful discussions. This work was supported by the National Natural Science Foundation (Grant No. 11704025).

\begin{widetext}

\section{Appendix A: The expressions for the partition function and $\langle \hat{n}^{\gamma}\rangle$}\label{app:appa}

In this appendix, we analytically derive the partition function of the DM as well as some expected values of observables by employing the method reported in Refs.~\cite{PhysRevA.9.418,PhysRevA.70.033808,Liberti2005,Bastarrachea_Magnani_2016,PhysRevE.96.012121}. The partition function of $\hat{H}_{\mathrm{D}}$ is defined by
\setcounter{equation}{0}
\renewcommand\theequation{A\arabic{equation}}
\begin{equation}
\mathcal{Z}_{\mathrm{D}}\equiv\mathrm{Tr}_{\mathrm{D}}\Big{(}e^{-\beta \hat{H}_{\mathrm{D}}}\Big{)}=\mathrm{Tr}_{\mathrm{a}}\mathrm{Tr}_{\mathrm{c}}\Big{(}e^{-\beta \hat{H}_{\mathrm{D}}}\Big{)},
\end{equation}
where
\begin{equation}
\mathrm{Tr}_{\mathrm{a}}\mathcal{\hat{O}}\equiv\sum_{\sigma_{1}=\mathrm{e},\mathrm{g}}\sum_{\sigma_{2}=\mathrm{e},\mathrm{g}}...\sum_{\sigma_{N}=\mathrm{e},\mathrm{g}}\langle \sigma_{1},\sigma_{2},...,\sigma_{N}|\mathcal{\hat{O}}|\sigma_{1},\sigma_{2},...,\sigma_{N}\rangle,
\end{equation}
denotes the partial trace operator with respect to the degrees of freedom of the atoms and
\begin{equation}
\mathrm{Tr}_{\mathrm{c}}\mathcal{\hat{O}}\equiv\int_{-\infty}^{\infty}\frac{d^{2}\alpha}{\pi}\langle\alpha|\mathcal{\hat{O}}|\alpha\rangle,
\end{equation}
is the partial trace operator with respect to the degree of freedom of the cavity field. Here, $|\alpha\rangle$ denotes the coherent state, defined as $\hat{a}|\alpha\rangle=\alpha|\alpha\rangle$.
Then, the partition function can be approximately computed as follows:
\begin{equation}
\begin{split}
\mathcal{Z}_{\mathrm{D}}=&\mathrm{Tr}_{\mathrm{a}}\int_{-\infty}^{\infty}\frac{d^{2}\alpha}{\pi}\langle\alpha|e^{-\beta \hat{H}_{\mathrm{D}}}|\alpha\rangle\simeq\mathrm{Tr}_{\mathrm{a}}\int_{-\infty}^{\infty}\frac{d^{2}\alpha}{\pi}\mathrm{Tr}_{\mathrm{a}}\Big{(}e^{-\beta \langle\alpha|\hat{H}_{\mathrm{D}}|\alpha\rangle}\Big{)}\\
=&\int_{-\infty}^{\infty}\frac{d^{2}\alpha}{\pi}e^{-\beta\omega|\alpha|^{2}}\mathrm{Tr}_{\mathrm{a}}\exp\bigg{[}-\beta\bigg{(}\hat{J}_{z}+\frac{2g\mathrm{Re}\alpha}{\sqrt{N}}\hat{J}_{x}\bigg{)}\bigg{]}\\
=&\int_{-\infty}^{\infty}\frac{d^{2}\alpha}{\pi}e^{-\beta\omega|\alpha|^{2}}\Bigg{\{}2\cosh\Bigg{[}\beta\sqrt{\frac{\epsilon^{2}}{4}+\frac{4g^{2}(\mathrm{Re}\alpha)^{2}}{N}}\Bigg{]}\Bigg{\}}^{N},
\end{split}
\end{equation}
where we have used the following approximation $\langle\alpha|e^{-\beta \hat{H}_{\mathrm{D}}}|\alpha\rangle\simeq1-\beta\langle\alpha| \hat{H}_{\mathrm{D}}|\alpha\rangle\simeq e^{-\beta\langle\alpha|\hat{H}_{\mathrm{D}}|\alpha\rangle}$. Such an approximation is acceptable in the high-temperature regime. To handle the $d^{2}\alpha$-integral, we introduce $x\equiv\mathrm{Re}\alpha$ and $y\equiv\mathrm{Im}\alpha$, which results in $d^{2}\alpha=dxdy$ and $|\alpha|^{2}=x^{2}+y^{2}$. By doing so, the $y$-part of the integral can be immediately carried out, and then one can find
\begin{equation}
\begin{split}
\mathcal{Z}_{\mathrm{D}}=&\frac{1}{\sqrt{\pi\beta\omega}}\int_{-\infty}^{\infty}dxe^{-\beta\omega x^{2}}\Bigg{[}2\cosh\Bigg{(}\beta\sqrt{\frac{\epsilon^{2}}{4}+\frac{4g^{2}x^{2}}{N}}\Bigg{)}\Bigg{]}^{N}.
\end{split}
\end{equation}
The above expression is still intricate. We use the steepest descent method or Laplace's integral method to further simplify the above expression~\cite{PhysRevA.9.418,PhysRevA.70.033808,Liberti2005,Bastarrachea_Magnani_2016,PhysRevE.96.012121}. To this aim, we need to introduce a new variable, $z$, defined as $z\equiv x/\sqrt{N}$; then the expression of $\mathcal{Z}_{\mathrm{D}}$ can be rewritten as
\begin{equation}\label{eq:eqa6}
\begin{split}
\mathcal{Z}_{\mathrm{D}}=&\sqrt{\frac{N}{\pi\beta\omega}}\int_{-\infty}^{\infty}dze^{N\Phi(z)},
\end{split}
\end{equation}
where $\Phi(z)$ is given by Eq.~(\ref{eq:eq3}) in the main text. The form of the partition function in Eq.~(\ref{eq:eqa6}) is especially suitable for Laplace's integral method~\cite{PhysRevA.9.418}, which consists of approximating the exponential integrand around the
maximum of the function $\Phi(z)$. By employing the Laplace approximation, one can obtain
\begin{equation}\label{eq:eqa7}
\begin{split}
\mathcal{Z}_{\mathrm{D}}\simeq\sqrt{\frac{2}{\beta\omega\varphi(z_{0})}}e^{N\Phi(z_{0})},
\end{split}
\end{equation}
where $z_{0}$ is determined by $\phi(z_{0})=0$ and $\varphi(z)\equiv|\partial_{z}^{2}\Phi(z)|$. With Eq.~(\ref{eq:eqa7}) at hand, one can immediately derive the expression of the free energy, i.e., Eq.~(\ref{eq:eq2}) in the main text. And the expected value of $\hat{J}_{z}$ per atom is given by
\begin{equation}
j_{z}\equiv\frac{\langle \hat{J}_{z}\rangle}{N}=-\frac{1}{N\beta}\frac{\partial}{\partial\epsilon}\ln\mathcal{Z}_{\mathrm{D}}=-\frac{\epsilon}{2\sqrt{\epsilon^{2}+16g^{2}z_{0}^{2}}}\tanh\bigg{(}\frac{1}{2}\beta\sqrt{\epsilon^{2}+16g^{2}z_{0}^{2}}\bigg{)}.
\end{equation}

Next, we derive the expression of $\langle \hat{n}^{\gamma}\rangle$ by making use of Laplace's integral method. In the high-temperature regime, one finds the mean photon number can be approximately derived as follows:
\begin{equation}
\begin{split}
\langle \hat{n}^{\gamma}\rangle=&\frac{1}{\mathcal{Z}_{\mathrm{D}}}\mathrm{Tr}_{\mathrm{D}}\Big{(}e^{-\beta \hat{H}_{\mathrm{D}}}\hat{n}^{\gamma}\Big{)}\simeq\frac{1}{\mathcal{Z}_{\mathrm{D}}}\mathrm{Tr}_{\mathrm{D}}\Big{(}e^{-\beta \hat{H}_{\mathrm{D}}+\ln \hat{n}^{\gamma}}\Big{)}\\
=&\frac{1}{\mathcal{Z}_{\mathrm{D}}}\int\frac{d^{2}\alpha}{\pi}\langle\alpha|e^{-\beta \hat{H}_{\mathrm{D}}+\ln \hat{n}^{\gamma}}|\alpha\rangle\simeq\frac{1}{\mathcal{Z}_{\mathrm{D}}}\int\frac{d^{2}\alpha}{\pi}e^{-\beta \langle\alpha|\hat{H}_{\mathrm{D}}|\alpha\rangle+\ln \langle\alpha|\hat{n}^{\gamma}|\alpha\rangle}\\
=&\frac{1}{\mathcal{Z}_{\mathrm{D}}}\int\frac{d^{2}\alpha}{\pi}|\alpha|^{2\gamma}e^{-\beta\omega|\alpha|^{2}}\Bigg{\{}2\cosh\Bigg{[}\beta\sqrt{\frac{\epsilon^{2}}{4}+\frac{4g^{2}(\mathrm{Re}\alpha)^{2}}{N}}\Bigg{]}\Bigg{\}}^{N}.
\end{split}
\end{equation}
Let $x=\mathrm{Re}\alpha$ and $y=\mathrm{Im}\alpha$; one can find
\begin{equation}
\begin{split}
\langle \hat{n}^{\gamma}\rangle=&\frac{1}{\pi}\frac{1}{\mathcal{Z}_{\mathrm{D}}}\int_{-\infty}^{\infty} dx\int_{-\infty}^{\infty} dy(x^{2}+y^{2})^{\gamma}\exp\Bigg{\{}-\beta\omega(x^{2}+y^{2})+N\ln\Bigg{[}2\cosh\Bigg{(}\beta\sqrt{\frac{\epsilon^{2}}{4}+\frac{4g^{2}x^{2}}{N}}\Bigg{)}\Bigg{]}\Bigg{\}}\\
=&\frac{1}{\pi}\frac{1}{\mathcal{Z}_{\mathrm{D}}}\int_{-\infty}^{\infty} dx\int_{-\infty}^{\infty} dy \sum_{k=0}^{\gamma}C_{k}^{\gamma}
(x^2)^{\gamma-k}y^{2k}\exp\Bigg{\{}-\beta\omega(x^{2}+y^{2})+N\ln\Bigg{[}2\cosh\Bigg{(}\beta\sqrt{\frac{\epsilon^{2}}{4}+\frac{4g^{2}x^{2}}{N}}\Bigg{)}\Bigg{]}\Bigg{\}},
\end{split}
\end{equation}
where $C_{k}^{\gamma}\equiv\frac{\gamma !}{k !(\gamma-k)!}$ is the binomial coefficient. The integral of variable $y$ can be exactly carried out using the Gaussian integral formula. Then, by introducing $z\equiv x/\sqrt{N}$ and applying the Laplace approximation again, we finally have
\begin{equation}
\begin{split}
\langle \hat{n}^{\gamma}\rangle=&\frac{1}{\pi}\frac{1}{\mathcal{Z}_{\mathrm{D}}}\sum_{k=0}^{\gamma}C_{k}^{\gamma}\frac{\Gamma{(k+\frac{1}{2})}}{{(\beta\omega)}^{k+\frac{1}{2}}}N^{\gamma-k+\frac{1}{2}}\int_{-\infty}^{\infty} dz{(z^{2})^{\gamma-k}}e^{N\Phi(z)}\\
\simeq&\frac{1}{\pi}\frac{1}{\mathcal{Z}_{\mathrm{D}}}\sum_{k=0}^{\gamma}C_{k}^{\gamma}\frac{\Gamma{(k+\frac{1}{2})}}{{(\beta\omega)}^{k+\frac{1}{2}}}N^{\gamma-k+\frac{1}{2}}{(z_{0}^{2})^{\gamma-k}}\sqrt{\frac{2\pi}{N\varphi(z_{0})}}e^{N\Phi(z_{0})}\\
=&\frac{1}{\sqrt\pi}\sum_{k=0}^{\gamma}C_{k}^{\gamma}\frac{\Gamma{(k+\frac{1}{2})}}{{(\beta\omega)}^{k}}{(z_{0}^{2})^{\gamma-k}}N^{\gamma-k},
\end{split}
\end{equation}
where $\Gamma(z)$ is Euler's gamma function.

\section{Appendix B: The effective Hamiltonian}\label{app:appb}

In our sensing scheme, a qubit is utilized as the quantum sensor to reveal the parameter $g$ in the DM. The Hamiltonian of the probe-qubit can be described by $\hat{H}_{\mathrm{s}}=\omega_{\mathrm{q}}\hat{\sigma}_{+}\hat{\sigma}_{-}$, where $\omega_{\mathrm{q}}$ denotes the frequency of the qubit. The probe-qubit is injected into the single-mode cavity and consequently interacts with the cavity field~\cite{PhysRevA.80.063829,PhysRevA.87.024101,Wu2016}. We assume the sensor-cavity interaction Hamiltonian is described as
\setcounter{equation}{0}
\renewcommand\theequation{B\arabic{equation}}
\begin{equation}
\begin{split}
\hat{H}_{\mathrm{qc}}=g_{\mathrm{qc}}\Big{(}\hat{\sigma}_{+}\hat{a}+\hat{\sigma}_{-}\hat{a}^{\dagger}\Big{)},
\end{split}
\end{equation}
where we have made the rotating wave approximation. If the probe-qubit is far-off-resonant with the cavity field, namely, $\Delta_{\mathrm{q}}\equiv\omega-\omega_{\mathrm{q}}\gg g_{\mathrm{qc}}$, one can use the Fr$\mathrm{\ddot{o}}$hlich-Nakajima transformation~\cite{PhysRev.79.845,doi:10.1080/00018735500101254}
\begin{equation}
\begin{split}
\hat{H}=e^{\mathcal{\hat{S}}}\Big{(}\hat{H}_{\mathrm{s}}+\hat{H}_{\mathrm{D}}+\hat{H}_{\mathrm{qc}}\Big{)}e^{-\mathcal{\hat{S}}},
\end{split}
\end{equation}
where $\mathcal{\hat{S}}=\chi(\hat{\sigma}_{+}\hat{a}-\hat{\sigma}_{-}\hat{a}^{\dagger})$, with $\chi\equiv g_{\mathrm{qc}}/\Delta_{\mathrm{q}}$, to eliminate the unimportant higher order terms in $\hat{H}$. Such treatment was used in Ref.~\cite{PhysRevA.80.063829}and can be equivalently done by using the adiabatic elimination approach~\cite{PhysRevA.72.033818,PhysRevB.76.024517}. Up to the order of $\chi^{2}$, one can find
\begin{equation}
\begin{split}
e^{\mathcal{\hat{S}}}\hat{H}_{\mathrm{s}}e^{-\mathcal{\hat{S}}}\simeq \omega_{\mathrm{q}}\hat{\sigma}_{+}\hat{\sigma}_{-}-\omega_{\mathrm{q}}\chi(\hat{\sigma}_{-}\hat{a}^{\dagger}+\hat{\sigma}_{+}\hat{a})-\omega_{\mathrm{q}}\chi^{2}(\hat{\sigma}_{+}\hat{\sigma}_{-}-\hat{a}^{\dagger}\hat{a}+2\hat{\sigma}_{+}\hat{\sigma}_{-}\hat{a}^{\dagger}\hat{a}),
\end{split}
\end{equation}
\begin{equation}
\begin{split}
e^{\mathcal{\hat{S}}}\hat{H}_{\mathrm{D}}e^{-\mathcal{\hat{S}}}\simeq &\epsilon\hat{J}_{z}+\omega \hat{a}^{\dagger}\hat{a}+\omega\chi(\hat{\sigma}_{-}\hat{a}^{\dagger}+\hat{\sigma}_{+}\hat{a})+\omega\chi^{2}(\hat{\sigma}^{+}\hat{\sigma}^{-}+\hat{\sigma}_{z}\hat{a}^{\dagger}\hat{a})\\
&+\frac{g}{\sqrt{N}}(\hat{J}_{+}+\hat{J}_{-})(\hat{a}^{\dagger}+\hat{a})+\frac{g\chi}{\sqrt{N}}(\hat{J}_{+}+\hat{J}_{-})(\hat{\sigma}_{+}+\hat{\sigma}_{-})+\frac{g\chi^{2}}{2\sqrt{N}}(\hat{J}_{+}+\hat{J}_{-})(\hat{\sigma}_{z}\hat{a}^{\dagger}+\hat{\sigma}_{z}\hat{a}),
\end{split}
\end{equation}
\begin{equation}
\begin{split}
e^{\mathcal{\hat{S}}}\hat{H}_{\mathrm{qc}}e^{-\mathcal{\hat{S}}}\simeq g_{\mathrm{qc}}(\hat{\sigma}_{+}\hat{a}+\hat{\sigma}_{-}\hat{a}^{\dagger})+2g_{\mathrm{qc}}\chi(\hat{\sigma}_{+}\hat{\sigma}_{-}+\hat{\sigma}_{z}\hat{a}^{\dagger}\hat{a})-2g_{\mathrm{qc}}\chi^{2}(\hat{\sigma}_{-}\hat{a}^{\dagger}+\hat{\sigma}_{+}\hat{a}+\sigma_{-}\hat{a}^{\dagger}\hat{a}^{\dagger}\hat{a}+\hat{\sigma}_{+}\hat{a}^{\dagger}\hat{a}\hat{a}),
\end{split}
\end{equation}
In the regime where $\omega\gg \max\{\omega_{\mathrm{q}},g_{\mathrm{qc}}\}$ and $N\gg 1$, we finally find an effective Hamiltonian as
\begin{equation}
\begin{split}
\hat{H}_{\mathrm{eff}}\simeq&\Big{(}\omega_{\mathrm{q}}+\frac{3g_{\mathrm{qc}}^{2}}{\Delta_{\mathrm{q}}}\Big{)}\hat{\sigma}_{+}\hat{\sigma}_{-}+\hat{H}_{\mathrm{D}}+(\omega\chi^{2}+2g_{\mathrm{qc}}\chi-\omega_{\mathrm{q}}\chi^{2})\hat{\sigma}_{z}\hat{a}^{\dagger}\hat{a}.
\end{split}
\end{equation}
By defining the renormalized sensor frequency $\omega_{\mathrm{s}}\equiv\omega_{\mathrm{q}}+3g_{\mathrm{qc}}^{2}/\Delta_{\mathrm{q}}$ and renormalized senor-DM coupling strength $\lambda\equiv\omega\chi^{2}+2g_{\mathrm{qc}}\chi-\omega_{\mathrm{q}}\chi^{2}$, one can retrieve Eq.~(\ref{eq:eq8}) in the main text.

\section{Appendix C: The decoherence factor}\label{app:appc}

In this appendix, we show how to derive the expression for the decoherence factor $\mathcal{L}(t)$. When the coupling strength between the sensor and the DM is sufficiently weak, namely, $\lambda/\omega_{\mathrm{s}}\ll 1$, one can use the Baker-Campbell-Hausdorff theorem to reexpress $e^{-it\hat{H}_{\mathrm{e,g}}}$ as
\setcounter{equation}{0}
\renewcommand\theequation{C\arabic{equation}}
\begin{equation}
e^{-it\hat{H}_{\mathrm{e}}}=e^{-it(\hat{H}_{\mathrm{D}}+\lambda \hat{n}+\omega_{\mathrm{s}})}\simeq e^{-\frac{1}{2}it\hat{H}_{\mathrm{D}}}e^{-it\lambda \hat{n}}e^{\frac{1}{2}it\hat{H}_{\mathrm{D}}}e^{-it\omega_{\mathrm{s}}},
\end{equation}
\begin{equation}
e^{it\hat{H}_{\mathrm{g}}}=e^{it(\hat{H}_{\mathrm{D}}-\lambda \hat{n})}\simeq e^{\frac{1}{2}it\hat{H}_{\mathrm{D}}}e^{-it\lambda \hat{n}}e^{\frac{1}{2}it\hat{H}_{\mathrm{D}}}.
\end{equation}
Then, the decoherence factor $\mathcal{L}(t)$ can be approximately written as
\begin{equation}
\begin{split}
\mathcal{L}(t)\simeq&\frac{1}{\mathcal{Z}}e^{-it\omega_{\mathrm{s}}}\mathrm{Tr}_{\mathrm{D}}\Big{(}e^{-\beta\hat{H}_{\mathrm{D}}}e^{\frac{1}{2}it\hat{H}_{\mathrm{D}}}e^{it\lambda \hat{n}}e^{\frac{1}{2}it\hat{H}_{\mathrm{D}}}e^{-\frac{it}{2}\hat{H}_{\mathrm{D}}}e^{-it\lambda \hat{n}}e^{\frac{it}{2}\hat{H}_{\mathrm{D}}}\Big{)}\\
=&\frac{1}{\mathcal{Z}}e^{-it\omega_{\mathrm{s}}}\mathrm{Tr}_{\mathrm{D}}\Big{(}e^{-\beta\hat{H}_{\mathrm{D}}}e^{-2it\lambda \hat{n}}\Big{)}\simeq\frac{1}{\mathcal{Z}}e^{-it\omega_{\mathrm{s}}}\mathrm{Tr}_{\mathrm{D}}\Big{[}e^{-\beta\hat{H}_{\mathrm{D}}}(1-2it\lambda \hat{n}-2\lambda^{2}t^{2}\hat{n}^{2})\Big{]}\\
=&e^{-it\omega_{\mathrm{s}}}(1-2it\lambda\langle\hat{n}\rangle-2\lambda^{2}t^{2}\langle\hat{n}^{2}\rangle)\simeq e^{-it\omega_{\mathrm{s}}}e^{-2it\lambda\langle \hat{n}\rangle}e^{-\lambda^{2}t^{2}\langle \hat{n}^{2}\rangle},
\end{split}
\end{equation}
which recovers Eq.~(\ref{eq:eq11}) in the main text.

\section{Appendix D: Classical and quantum FIs}\label{app:appd}

For the sensing scheme considered in this paper, the output state of the sensor $\rho_{\mathrm{s}}(t)$ is a $2\times 2$ density matrix in the basis $\{|\mathrm{e}\rangle,|\mathrm{g}\rangle\}$; thus, the measurement operators can be constructed as $\{\mathcal{\hat{M}}_{\pm}=|\pm\rangle\langle\pm|\}$, where $|\pm\rangle=\frac{1}{\sqrt{2}}(|\mathrm{e}\rangle\pm|\mathrm{g}\rangle)$ with $|\pm\rangle$ being the eigenstates of $\hat{\sigma}_{x}$, i.e., $\hat{\sigma}_{x}|\pm\rangle=\pm|\pm\rangle$. The corresponding probability distributions for $\{|\pm\rangle\langle\pm|\}$ are given by
\setcounter{equation}{0}
\renewcommand\theequation{D\arabic{equation}}
\begin{equation}
p(\pm|g)=\frac{1}{2}\pm\frac{1}{2}\cos\big{(}\omega_{\mathrm{s}}t+2\lambda t\langle \hat{n}\rangle\big{)}e^{-\lambda^{2}t^{2}\langle\hat{n}^{2}\rangle}.
\end{equation}
Then, with the help of Eq.~(\ref{eq:eq6}), the classical FI is given by
\begin{equation}
\begin{split}
F_{g}(t)=&\frac{\lambda^{2}t^{2}e^{-2\lambda^{2}t^{2}\langle\hat{n}^{2}\rangle}}{1-\cos^{2}\big{(}\omega_{\mathrm{s}}t+2\lambda t\langle \hat{n}\rangle\big{)}e^{-2\lambda^{2}t^{2}\langle\hat{n}^{2}\rangle}}\Big{[}2(\partial_{g}\langle \hat{n}\rangle)\sin\big{(}\omega_{\mathrm{s}}t+2\lambda t\langle \hat{n}\rangle\big{)}+\lambda t(\partial_{g}\langle \hat{n}^{2}\rangle)\cos\big{(}\omega_{\mathrm{s}}t+2\lambda t\langle \hat{n}\rangle\big{)}\Big{]}^{2},
\end{split}
\end{equation}
where
\begin{equation}\label{eq:eqd3}
\partial_{g}\langle \hat{n}\rangle=N\partial_{g}(z_{0}^{2}),~~~\partial_{g}\langle \hat{n}^{2}\rangle=\frac{N}{\beta\omega}\partial_{g}(z_{0}^{2})+2N^{2}z_{0}^{2}\partial_{g}(z_{0}^{2}).
\end{equation}
The explicit expression of $\partial_{g}(z_{0}^{2})$ can be derived from Eq.~(\ref{eq:eq5}) and the result reads
\begin{equation}\label{eq:eqd4}
\partial_{g}(z_{0}^{2})=\frac{\epsilon^{2}}{4g^{3}}\bigg{[}\frac{\eta^{2}-1}{2}+\frac{\omega\eta^{2}}{\omega-2\beta g^{2}\mathrm{sech}^{2}(\beta\eta\epsilon/2)}\bigg{]}.
\end{equation}

On the other hand, one can rewrite $\rho_{\mathrm{s}}(t)$ in the Bloch representation in which the Bloch vector is given by $\pmb{\vec{r}}\equiv[\mathrm{Re}\mathcal{L}(t),-\mathrm{Im}\mathcal{L}(t),0]^{\mathbb{T}}$. Then, the quantum FI can be computed using Eq.~(\ref{eq:eq7}) in the main text, and we find
\begin{equation}\label{eq:eqd5}
\begin{split}
\mathcal{F}_{g}(t)=&\frac{1}{2}\lambda^{2}t^{2}\bigg{\{}8(\partial_{g}\langle \hat{n}\rangle)^{2}e^{-2\lambda^{2}t^{2}\langle \hat{n}^{2}\rangle}+\lambda^{2}t^{2}(\partial_{g}\langle \hat{n}^{2}\rangle)^{2}\Big{[}\coth(\lambda^{2}t^{2}\langle \hat{n}^{2}\rangle)-1\Big{]}\bigg{\}}.
\end{split}
\end{equation}
For the GHZ-state case, the quantum FI is given by
\begin{equation}
\begin{split}
\mathcal{F}_{g}^{\mathrm{GHZ}}(t)=&\frac{1}{2}\mathbb{N}^{2}\lambda^{2}t^{2}\bigg{\{}8(\partial_{g}\langle \hat{n}\rangle)^{2}e^{-2\mathbb{N}\lambda^{2}t^{2}\langle \hat{n}^{2}\rangle}+\lambda^{2}t^{2}(\partial_{g}\langle \hat{n}^{2}\rangle)^{2}\Big{[}\coth(\mathbb{N}\lambda^{2}t^{2}\langle \hat{n}^{2}\rangle)-1\Big{]}\bigg{\}},
\end{split}
\end{equation}
which reduces to Eq.~(\ref{eq:eqd5}) when $\mathbb{N}=1$.

\section{Appendix E: Multi-parameter sensing case}\label{app:appe}

\setcounter{equation}{0}
\renewcommand\theequation{E\arabic{equation}}
\setcounter{figure}{0}
\renewcommand\thefigure{A\arabic{figure}}
\begin{figure}
\centering
\includegraphics[angle=0,width=10.5cm]{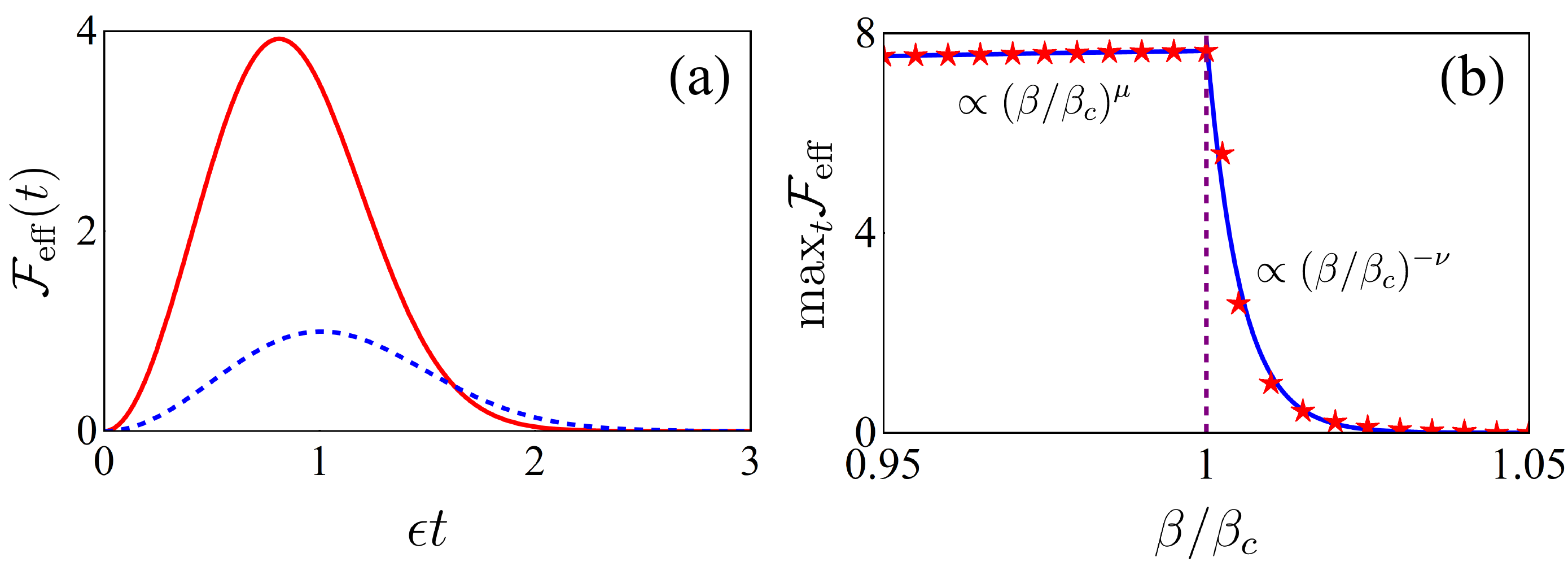}
\caption{(a) The effective quantum FI, $\mathcal{F}_{\mathrm{eff}}(t)$, is plotted as the function of the encoding time in different phases: $\beta/\beta_{c}=0.5$ (red solid line) and $\beta/\beta_{c}=1.01$ (blue dashed line). (b) The maximal effective quantum FI with respect to the optimal encoding time is plotted as the function of $\beta/\beta_{c}$. The red stars are numerical results, and the blue solid line is from the least-squares fitting. Other parameters are the same with these of Fig.~\ref{fig:fig1}.}\label{fig:fig6}
\end{figure}

In this appendix, we discuss the multi-parameter estimation case within the framework of our sensing scheme. To this aim, it is necessary to recall some basic formulation about the quantum FI matrix and the corresponding quantum Cram$\mathrm{\acute{e}}$r-Rao bound in quantum multi-parameter estimation theory.

Let us consider a vector of parameters $\pmb{\zeta}=(\zeta_{1},\zeta_{2},\zeta_{3},...,\zeta_{\mathrm{m}},...)^{\mathbb{T}}$, where $\zeta_{\mathrm{m}}$ is the $\mathrm{i}$th parameter of interest. The vector $\pmb{\zeta}$ is encoded into a quantum state $\rho(\pmb{\zeta})$. Then, by measuring $\rho(\pmb{\zeta})$, the message about $\pmb{\zeta}$ can be extracted from the outcomes. In such a multi-parameter estimation case, the estimation uncertainty, $\mathrm{cov}(\pmb{\zeta})$, which is defined by the covariance matrix of $\pmb{\zeta}$, is bounded by the following inequality~\cite{Liu_2019,PhysRevA.90.062113}:
\begin{equation}\label{eq:eqe1}
\mathrm{cov}(\pmb{\zeta})\geq \pmb{\mathcal{F}}(\pmb{\zeta})^{-1},
\end{equation}
which is the quantum Cram$\mathrm{\acute{e}}$r-Rao bound in the multi-parameter version. Here, $\pmb{\mathcal{F}}(\pmb{\zeta})^{-1}$ should be regarded as the matrix inverse of the quantum FI matrix $\pmb{\mathcal{F}}(\pmb{\zeta})$, which is defined by $\pmb{\mathcal{F}}_{\mathrm{mn}}(\pmb{\zeta})=\frac{1}{2}\mathrm{Tr}[\rho(\pmb{\zeta})\{\hat{\varsigma}_{\mathrm{m}},\hat{\varsigma}_{\mathrm{n}}\}]$, with $\hat{\varsigma}_{\mathrm{m}}$ determined by $\partial_{\zeta_{\mathrm{m}}}\rho(\pmb{\zeta})=\frac{1}{2}[\rho(\pmb{\zeta})\zeta_{\mathrm{m}}+\zeta_{\mathrm{m}}\rho(\pmb{\zeta})]$. Generally, some elements of $\pmb{\zeta}$ may be more important than others. Thus, a diagonal matrix $\pmb{\mathcal{W}}$, which is called the weighting matrix, is commonly introduced to define the following scalar quantity
\begin{equation}
\Theta(\pmb{\zeta})\equiv \mathrm{Tr}[\pmb{\mathcal{W}}\mathrm{cov}(\pmb{\zeta})]
\end{equation}
as the witness of the multi-parameter estimation precision~\cite{PhysRevA.90.062113,PhysRevLett.120.080501,PhysRevA.87.012107}. Together with Eq.~(\ref{eq:eqe1}), one can see the quantum multi-parameter Cram$\mathrm{\acute{e}}$r-Rao bound implies~\cite{PhysRevLett.120.080501}
\begin{equation}
\Theta(\pmb{\zeta})\geq\sum_{\mathrm{m}}\pmb{\mathcal{W}}_{\mathrm{mm}}[\pmb{\mathcal{F}}(\pmb{\zeta})^{-1}]_{\mathrm{mm}}.
\end{equation}
For the simplest two-parameter estimation case, one can easily find
\begin{equation}
\pmb{\mathcal{F}}(\pmb{\zeta})^{-1}=\frac{1}{\mathrm{Det}[\pmb{\mathcal{F}}(\pmb\zeta)]}\left[
                                                                                                            \begin{array}{cc}
                                                                                                              \mathcal{F}_{22}(\pmb\zeta) & -\mathcal{F}_{12}(\pmb\zeta) \\
                                                                                                              -\mathcal{F}_{12}(\pmb\zeta) & \mathcal{F}_{11}(\pmb\zeta) \\
                                                                                                            \end{array}
                                                                                                          \right],
\end{equation}
where $\mathrm{Det}$ denotes the determinant. By choosing $\pmb{\mathcal{W}}$ to be a $2\times2$ identity matrix, we have
\begin{equation}\label{eq:eqe5}
\Theta(\pmb{\zeta})\geq\frac{1}{\mathcal{F}_{\mathrm{eff}}}=\bigg{\{}\frac{\mathrm{Det}[\pmb{\mathcal{F}}(\pmb{\zeta})]}{\mathrm{Tr}[\pmb{\mathcal{F}}(\pmb{\zeta})]}\bigg{\}}^{-1}.
\end{equation}
Equation~(\ref{eq:eqe5}) has the same structure as that of the single-parameter estimation case. In this sense, $\mathcal{F}_{\mathrm{eff}}$ can be treated as effective quantum FI, which plays the same role as quantum FI, in the two-parameter estimation case~\cite{Liu_2019}.

Next, employing the above formulation, we discuss a two-parameter sensing case within the framework of our sensing scheme. The frequency of the cavity field $\omega$ and the atom-cavity coupling strength $g$ are the two parameters to be estimated; that is, we choose $\pmb{\zeta}=(\omega,g)^{\mathbb{T}}$. With the help of Eq.~(\ref{eq:eq9}), we find
\begin{equation}\label{eq:eqe6}
\begin{split}
\mathcal{F}_{\mathrm{mn}}(t)=&\frac{1}{2}\lambda^{2}t^{2}\bigg{\{}8(\partial_{\mathrm{m}}\langle \hat{n}\rangle)(\partial_{\mathrm{n}}\langle \hat{n}\rangle)e^{-2\lambda^{2}t^{2}\langle \hat{n}^{2}\rangle}+\lambda^{2}t^{2}(\partial_{\mathrm{m}}\langle \hat{n}^{2}\rangle)(\partial_{\mathrm{n}}\langle \hat{n}^{2}\rangle)\Big{[}\coth(\lambda^{2}t^{2}\langle \hat{n}^{2}\rangle)-1\Big{]}\bigg{\}},
\end{split}
\end{equation}
where $\partial_{g}\langle \hat{n}\rangle$ and $\partial_{g}\langle \hat{n}^{2}\rangle$ have already been given by Eqs.~(\ref{eq:eqd3}) and~(\ref{eq:eqd4}), while
\begin{equation}
\partial_{\omega}\langle \hat{n}\rangle=-\frac{1}{2\beta\omega^{2}}+N\partial_{\omega}(z_{0}^{2}),~~~\partial_{\omega}\langle \hat{n}^{2}\rangle=-\frac{3}{2\beta^{2}\omega^{3}}-\frac{N}{\beta\omega^{2}}z_{0}^{2}+\frac{N}{\beta\omega}\partial_{\omega}(z_{0}^{2})+2N^{2}z_{0}^{2}\partial_{\omega}(z_{0}^{2}),
\end{equation}
with
\begin{equation}
\partial_{\omega}(z_{0}^{2})=\frac{\epsilon^{2}\eta^{2}}{16\beta g^{4}\mathrm{sech}^{2}(\beta\eta\epsilon/2)-8g^{4}\omega}.
\end{equation}
With Eq.~(\ref{eq:eqe6}) at hand, the expression for the effective quantum FI can easily be obtained. In Fig.~\ref{fig:fig6}(a), we display the time evolution of $\mathcal{F}_{\mathrm{eff}}(t)$ in both normal and super-radiant phases. We can see $\mathcal{F}_{\mathrm{eff}}(t)$ has a dynamical behavior similar to that of the single-parameter sensing case. An optimal encoding time, which can maximize the value of $\mathcal{F}_{\mathrm{eff}}(t)$, is found. In Fig.~\ref{fig:fig6}(b), we plot the maximal effective quantum FI with respect to the optimal encoding time as a function of $\beta/\beta_{c}$. A peak occurs at the thermodynamic critical point $\beta/\beta_{c}=1$. This result means our thermodynamic-criticality-enhanced scheme can be generalized to the multi-parameter sensing case without difficulties.

\end{widetext}
	
\bibliography{reference}

\end{document}